\documentstyle[aps,prb,preprint,epsf,floats]{revtex}

\begin{document}

 \bibliographystyle{prsty}


 \title{
 Oscillation of the tunnel splitting in nanospin systems within the particle mapping formalism}

 \author{Sahng-Kyoon Yoo$^{a,c}$, Soo-Young Lee$^b$, Dal-Ho Yoon$^c$ and Chang-Soo Park$^d$}

 \address{$^a$ Department of Physics, Seonam University, Namwon, 590-711, Korea}

 \address{$^b$ Department of  Physics, Kyungnam University, Masan 631-701, Korea}

 \address{$^c$ Department of Physics, Chongju University, Chongju 360-764, Korea}

 \address{$^d$ Department of Physics, Dankook University, Cheonan 330-714, Korea\\
\smallskip
{\rm (\today)}
\bigskip\\
\parbox{14.2 cm}
{\rm
 The oscillation of tunnel splitting in the biaxial spin systems with the magnetic field
 along the hard anisotropy axis is analyzed within the particle mapping apporach, rather than in the
 $(\theta, \phi)$ spin coherent-state representation.
 In our mapping procedure, the spin system is transformed into a particle moving in the restricted $S^1$
 geometry whose wave function subjects to the boundary condition involving additional phase shift.
 We obtain the new topological phase that plays the same
 role as the Wess-Zumino action in spin coherent-state representation.
 Considering the interference of two possible trajectories, instanton and anti-instanton,
 we get the identical condition for the field at which tunneling is quenched,
 with the previous result within spin coherent-state representation.
\smallskip
 \begin{flushleft}
 \end{flushleft}
}}

\maketitle

\newpage

 For the last decade, there have been a lot of works on the macroscopic magnetization reversal in nanomagnetic
 single-domain particles with many spins both theoretically\cite{chudnovsky88,gunther95} and
 experimentally\cite{gunther95,awschalom90,friedman96,sangregorio97}. It is well-known that,
 at low temperature, the rate of such reversal is dominated by quantum tunneling through the barrier, while
 at high temperatures, it is governed by classical thermal activation according to the law $\exp (-E_0 / k_B T)$,
 where $E_0$ is the barrier energy and $T$ is the temperature. Moreover, two kinds of tunneling mechanism
 exist in tunneling regime, the macroscopic quantum tunneling (MQT) from a metastable state to a stable one,
 and the macroscopic quantum coherence (MQC) between the two equivalent ground states. These have been
 treated by spin coherent-state path integral formalism\cite{liang98}
 or direct mapping onto a particle\cite{scharf87,ulyanov92,chudnovsky97,park99} in uniaxial and biaxial spin systems.

 Here, we focus on the MQC of biaxial spin system like Fe$_8$ which shows the exotic phenomena. In the absence of
 magnetic field, Loss et. al.\cite{loss92} and Delft et. al.\cite{delft92} have shown that the tunneling is quenched
 for half-integer spins, and
 allowed for integer spins using the coherent state path integral. Such spin parity effect is orginated from the
 topological Wess-Zumino phase in Euclidean action in path integral\cite{fradkin91}
 and is directly related to the well-known
 Kramer's degeneracy. Later, Garg\cite{garg93} first have shown,
 also within the coherent-state path integral, that in the presence of magnetic field
 along the hard anisotropy axis, i.e., in the system decribed by Hamiltonian
 \begin{equation}
 {\cal H}= k_1 S_z^2 + k_2 S_x^2 - H S_z,
 \end{equation}
 where $k_1$ and $k_2$ are positive anisotropy constant with $k_1 > k_2$,
 the tunneling spilitting also vanishes when the external field $H$ takes the values
 \begin{equation}
 \label{vanishfield}
 H=2 k_2 \sqrt{1- \lambda} (S - n - 1/2 ),
 \end{equation}
 where $n=0, 1, \cdots, 2S-1$, $\lambda= k_2 /k_1$
 and $S$ is the total spin. Very recently, this system has been conformed by experiment\cite{wernsdorfer99},
 and also analyzed in more detail\cite{chudnovsky99}.
 These oscillations of level splittings are also due to the topological phase factor in spin coherent-state path
 integrals. These seemingly simple level crossing can be interpreted as the interference of two different
 classical trajectories in Euclidean space.

 Another approach, the direct mapping onto a particle, is also useful in that this
 enables us to treat the spin system more quantum mechanically, as
 well as gives a constant mass in the biaxial spin system.
 In this paper, we derive Eq. (\ref{vanishfield}) using the direct particle mapping rather than coherent-state
 path integral. In this procedure, the boundary condition of particle wave function in the presence of external
 field along the hard axis is modified by a topological phase shift in comparison to the zero-field case.
 We can interpret that a particle moves on a restricted $S^1$ geometry with wave function having not the periodic
 boundary condition but the additional geometrical phase. From this boundary condition, we can derive the
 overlap integral which produces the classical action in Euclidean space. Considering the interference of two
 possible trajactories, instanton and anti-instanton, we find the field values at which the tunneling vanishes.
 These values are just the same as the previous results obtained within the spin coherent-state path integral
 formalism.

 Consider first the simple uniaxial spin system which has hard axis anisotropy with magnetic field applied along the
 hard axis, in order to get an idea in a simpler system. The corresponding spin Hamiltonian is expressed as:
 \begin{equation}
 \label{uniham}
 {\cal H}= S_z^2 - H S_z.
 \end{equation}
 Note that this system does not correspond to tunneling.
 This Hamiltonian has eigenvalues $\varepsilon_m = m^2 - Hm$ in $S_z$-representation,
 where $m=-S, -S-1, \cdots, S-1, S$, and
 $S$ is the total spin number. The eigenvalues have many level crossings as the field increases.
 Although we already knew all the energy levels, for the pedagogical purpose,
 we perform the mapping onto a particle by defining the characteristic function
 \begin{equation}
 \label{unichar}
 \Phi (\phi) = \exp( i m \phi ).
 \end{equation}
 Following the well-known mapping procedure, we obtain the Schr\"{o}dinger-like equation
 \begin{equation}
 - \frac{d^2 \Psi}{d \phi^2} + \frac{H^2}{4} \Psi = \varepsilon \Psi,
 \end{equation}
 where
 \begin{equation}
 \label{uniwf}
 \Psi(\phi) = \Phi(\phi) \exp \left( \frac{iH}{2} \phi \right),
 \end{equation}
 The wave function $\Psi$ satisfies the boundary condition for $2 \pi$-rotation
 \begin{equation}
 \label{unibc}
 \Psi(\phi + 2\pi) = \pm \Psi(\phi) \exp ( i \pi H)
 \end{equation}
 where plus sign corresponds to bosons, while minus sign to fermions.
 This equation corresponds to the motion of particle with mass $m=1/2$ moving in the potential $V(\phi)=
 H^2 /4$ on the constrained geometry $S^1$.
 In above example, the phase factor appearing in the wave function, Eq. (\ref{uniwf}), plays a crucial role
 in yielding the original eigenvalues, $\varepsilon_m =m^2 - Hm$.
 Although the level crossing is merely the quantum mechanical
 property in spin systems, such effect can be interpreted as the
 result of
 the geometrical phase in the wave function of particle.

 Another uniaxial spin system, Mn$_{12}$Ac, whose Hamiltonian is given by ${\cal H}=-S_z^2 - H S_z$ can be
 considered. Following the above procedure, we obtain the equation describing the particle with negative mass.
 This corresponds to the motion in the inverted potential having $(2S+1)$-eigenvalues
 $- \varepsilon_m$ but in opposite order
 to those of original spin system. The level crossing in this case can also be shown by the same method.
 These observations are to be extended to the biaxial
 spin systems with external magnetic field applied along the hard anisotropic axis.

 Let's consider the biaxial spin Hamiltonian with the magnetic field in the hard $z$-direction
 \begin{equation}
 \label{biham}
 {\cal H} = - D S_x^2 + E S_z^2 - H S_z,
 \end{equation}
 where $D$ and $E$ are easy- and hard-axis anisotropy constants satisfying $D, E >0$.
 This system also can be mapped onto a particle system using the usual mapping procedure\cite{scharf87}.
 We define the characteristic function
 \begin{equation}
 \Phi (\phi) = \sum_{m=-S}^{S} \frac{c_m e^{im \phi}}{\sqrt{(S-m)! (S+m)!}},
 \end{equation}
 where $c_m$ is the wave function of spin system. Then, we obtain the differential equation for $\Phi$:
 \begin{equation}
 \label{bieqphi}
 - P(\phi)  \frac{d^2 \Phi}{d \phi^2} + Q(\phi) \frac{d \Phi}{d \phi} + R(\phi) \Phi = \varepsilon \Phi,
 \end{equation}
 where
 \begin{eqnarray}
 P(\phi) & = & 1 - \lambda + \lambda \sin^2 \phi , \nonumber \\
 Q(\phi) & = & 2 iSh + \lambda \left( S- \frac{1}{2} \right) \sin 2 \phi, \\
 R(\phi) & = & - \frac{\lambda}{2} (S^2 +S) - \frac{\lambda}{2} (S^2 -S) \cos 2 \phi, \nonumber
 \end{eqnarray}
 with $\lambda= D / (D+E) ,~ h=H/2 (D +E) S$, and $\varepsilon$ is the reduced energy
 eigenvalue. To eliminate the first-order derivative of $\Phi$ and make the particle mass constant,
 we can introduce the new wave function
 \begin{equation}
 \Psi(\phi)=\Phi(\phi) \exp( -G(\phi)),
 \end{equation}
 and change the variable from $\phi$ to $x=x(\phi)$, so that the Eq. (\ref{bieqphi}) can be transformed as
 \begin{equation}
 - \frac{d^2 \Psi}{dx^2} +V(x) \Psi = \varepsilon \Psi,
 \end{equation}
 where
 \begin{equation}
 \label{snx}
 x = \int^{\phi} \frac{d \theta}{\sqrt{1- \lambda + \lambda \sin^2 \theta}} = {\rm sn} ^{-1} (\sin \phi, \sqrt{\lambda}),
 \end{equation}
 Here, ${\rm sn} (x,k)$ is the usual Jacobian Elliptic funtion with period of $4 {\cal K} (k)$, where ${\cal K}$ is the
 complete elliptic function of first kind.
 Due to Eq. (\ref{snx}), $\sin \phi$ and $\cos \phi$ can be changed into ${\rm sn} x$ and ${\rm cn} x$, respectively.
 The resulting effective potential and wave function are given by
 \begin{eqnarray}
 V(x) & = &   \frac{1}{1- \lambda + \lambda {\rm sn}^2 x} \Bigg(
                   \lambda^2 S (S+1) {\rm {\rm sn}}^2 x {\rm cn}^2 x - h^2 S^2 \nonumber \\
        & &     + 2i \lambda hS \left( S+ \frac{1}{2} \right) {\rm sn} x {\rm cn} x \Bigg) - \lambda (S^2 +S) {\rm cn}^2 x
                   \nonumber \\ \\
 \Psi(x) & = & \Phi(x) (1- \lambda + \lambda {\rm sn}^2 x)^{-\frac{S}{2}} \nonumber \\
        & & \times \exp \left( - iSh \int^{x} \frac{{\rm dn} x' dx'}{1-\lambda + \lambda {\rm sn}^2 x'} \right)
 \end{eqnarray}
 Note that the effective particle potential corresponds to the complex potential.
 The real part of $V(x)$ is the periodic potential with period $4{\cal K}$.
 The imaginary part of $V(x)$ is related to the topological phase shift in wave function in the manner of Aharonov-Bohm
 effect, but does not contribute to the tunneling rate, which will be clear later. Since the integral
 \begin{equation}
 \int_{x}^{x+ 4 {\cal K} } \frac{{\rm dn} x' dx'} {1- \lambda + \lambda {\rm sn}^2 x'} = \frac{2 \pi} {\sqrt{1-\lambda}},
 \end{equation}
 wave function $\Psi$ satiesfies the following boundary condition:
 \begin{equation}
 \label{bibc}
 \Psi (x + 4 {\cal K}) = \Psi (x) \exp \left[ 2 \pi i S \left( 1- \frac{h}{\sqrt{1-\lambda}} \right) \right].
 \end{equation}

 Now, we consider the path integration in Euclidean space in order to get the tunneling rate. The amplitude from
 the state $\mid x>$ at Euclidean time $\tau=\tau_0$ to the state $\mid x'>$ at $\tau$ is
 \begin{equation}
 \label{biamplitude}
 <x', \tau \mid x, \tau_0 >
 = \int {\cal D} [x(\tau)] \prod_{k=1}^{N} <x_{k} \mid x_{k-1}>
 \exp \left( - \int_{\tau_0}^{\tau} d \tau {\cal H} \right),
 \end{equation}
 where $x_0 =x$ and $x_N =x'$.
 The overlap integral  between $k$th and $(k-1)$th space elements can be written in small $x_k - x_{k-1}$ limit as
 \begin{equation}
 < x_k \mid x_{k-1} > =
 \exp \left[ -i \frac{2 \pi S}{4 {\cal K}} (x_k - x_{k-1}) \left(1- \frac{h}{\sqrt{1-\lambda}} \right) \right].
 \end{equation}
 Therefore, the product of overlap integral between $k$th and $(k-1)$th space elements in Eq. (\ref{biamplitude})
 is given by
 \begin{eqnarray}
 \prod_{k=1}^{N} <x_k \mid x_{k-1}>
 & = & \exp \left[ -i \frac{\pi S}{2 {\cal K}} \sum_{k} \epsilon \dot{x}
          \left( 1- \frac{h}{\sqrt{1-\lambda}} \right) \right] \nonumber \\
 & = & \exp \left[ -i \frac{\pi S}{2 {\cal K}} \int_{\tau_0}^{\tau} d \tau \dot{x}
          \left( 1- \frac{h}{\sqrt{1-\lambda}} \right) \right],
 \end{eqnarray}
 where $\epsilon = \tau_k - \tau_{k-1}$.
 The imaginary-time effective action which gives the tunneling rate is therefore written as
 \begin{equation}
 \label{biaction}
 S_E = -i \frac{\pi S}{2 {\cal K}} \left( 1- \frac{h}{\sqrt{1-\lambda}} \right) \int dx
 + \int d \tau [ 4 \dot{x}^2 + V(x)].
 \end{equation}

 We consider the interferences of two possible trajectories of instanton and anti-instanton
 which move from $x=0$ to $x=2 {\cal K}$ and to $-2 {\cal K}$ in the real part of inverted potential.
 In this case, the real parts of tunneling rate are factored out and we only focus on the imaginary part of
 effective action in Eq. (\ref{biaction}):
 \begin{equation}
 {\rm Im} S_E =  - \frac{\pi S}{2 {\cal K}} \left( 1- \frac{h}{\sqrt{1-\lambda}} \right) \int dx +
 \int d \tau {\rm Im} V(x).
 \end{equation}
 Here, the contribution of second integral, that is, imaginary potential, is zero, due to its oddness.
 Therefore, we obtain the interferences of two trajectories which vanishes when
 \begin{equation}
 h=\sqrt{1-\lambda} \frac{S-n - 1/2}{S}, ~n=0, 1, \cdots, 2S-1.
 \end{equation}
 which is exactly same as those in the previous calculations within $(\theta, \phi)$ spin coherent-states
 representation\cite{garg93,chudnovsky99}

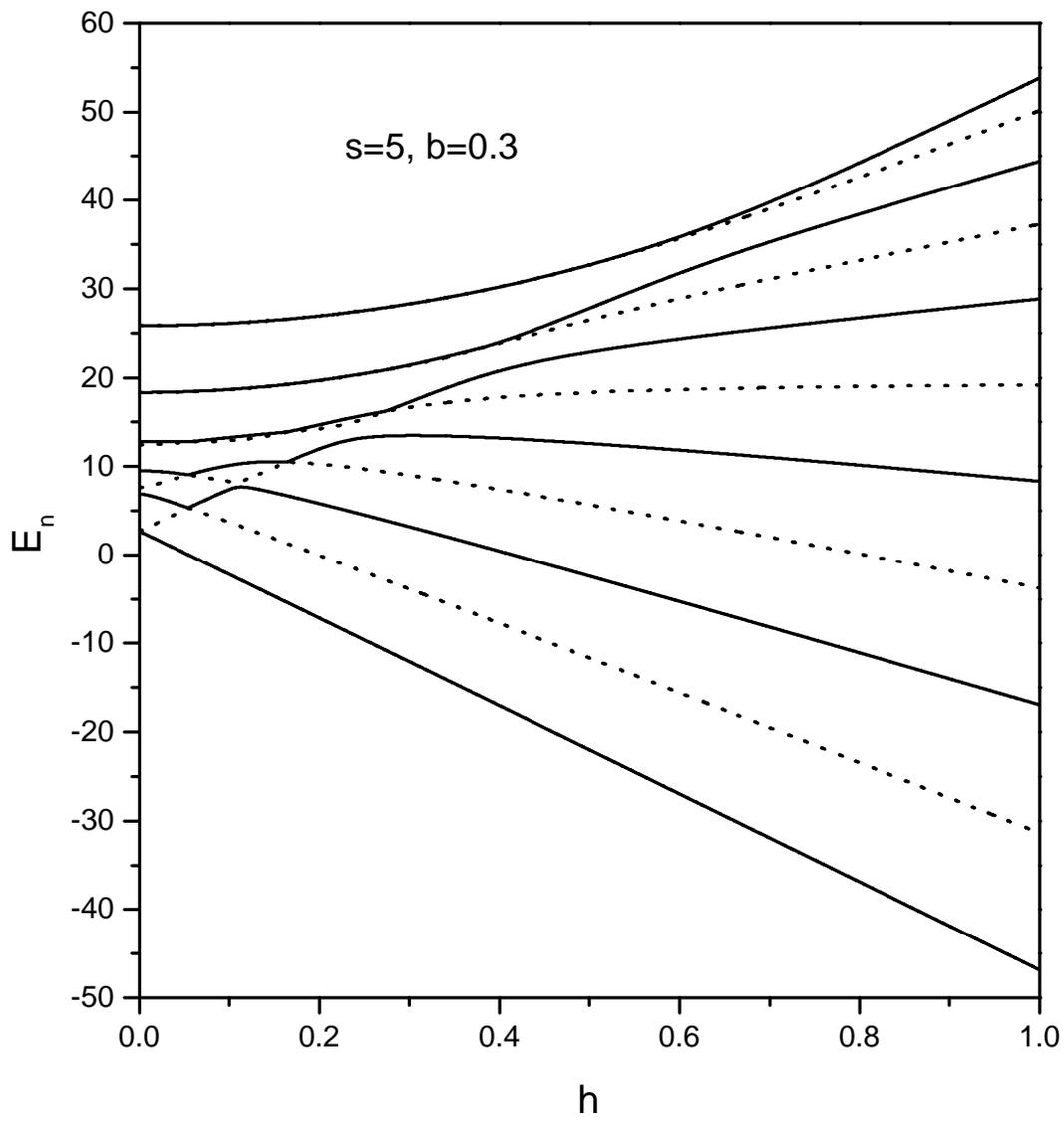
 \begin{figure}
 \caption{The lowest $(2S+1)$ energy eigenvalues of Eq. (\protect{\ref{bisch1}}).
 This structure is exactly same as that of original spin system expressed as Hamiltonian,
 Eq. (\protect{\ref{biham1}}) in inverted order. The points where
 the solid and dotted lines meet represent the level crossing.
 }
 \label{figure1}
 \end{figure}

 Here, we note that the variable $\phi$, and therefore $x$ seems be an azimuthal angle.
 When we choose the Hamiltonian as Eq.(\ref{biham}) among the possible Hamiltonians that have same physics,
 $\phi$ can be interpreted as the usual azimuthal angle in the $S_z$-representation, since $z$-axis is a hard axis.
 But in this case, while the particle mass is positive, the effective potential
 becomes complex, and therefore, the energy eigenvalues are also complex.
 However, the boundary condition with the additional phase, Eq. (\ref{bibc})  just exactly singles out
 the real energy eigenvalues\cite{ulyanov92}.
 We can choose the alternative Hamiltonian other than Eq. (\ref{biham}) in order for the
 imaginary potential not to appear in Schr\"{o}dinger equation. For example, let us consider the Hamiltonian
 \begin{equation}
 \label{biham1}
 {\cal H} = -AS_z^2 - BS_x^2 - H S_y,
 \end{equation}
 where $A$ and $B$ are positive constants with $A>B>0$. We can also make a direct mapping onto a particle
 as above. In this case, the variable $\phi$ in the characteristic function does not play a role as an
 azimuthal angle, so that the tunneling picture in original spin system does not have one-to-one correspondence
 with an effective particle system\cite{add1}. However, the effective eigenvalue equation is given by
 \begin{equation}
 \label{bisch1}
 \frac{d^2 \Psi (x)}{dx^2} - V(x) \Psi (x) = \epsilon \Psi (x),
 \end{equation}
 where
 \begin{eqnarray}
 \label{bipot1}
 V(x) & = & \frac{{\rm cn}^2 x}{{\rm sn}^2 x} \left[ b^2 S (S+1) {\rm sn}^2 x - 2bhS \left( S+ \frac{1}{2} \right)
 {\rm sn} x + h^2 S^2 \right]  \nonumber \\
 & & + bS (S+1) {\rm cn}^2 x + hS (2S+1) {\rm sn} x,
 \end{eqnarray}
 with $b \equiv B/A$ and $h \equiv 2 AS$. The boundary condition of particel wave function is given
 \begin{equation}
 \Psi(x + 4{\cal K}) = \pm \Psi (x).
 \end{equation}
 As shown above, the Hamiltonian gives the real potential with negative mass. Furthermore, the field-dependent
 additional geometrical phase in the boundary condition of wave function does not appear by this Hamiltonian.
 As discussed in the uniaxial case,
 this can be regarded as the motion in the inverted potential with eigenvalues having opposite sign and in
 opposite order to the original spin system. Through the simple numerical calculation, we can obtain the
 $(2S+1)$ inverted energy eigenvalues, which show the oscillation of tunnel splitting with external field.
 This is represented in Figure. \ref{figure1}

 In conclusion, in this paper we consider the oscillation of tunnel splitting in spin system within
 direct particle mapping approach. In simple uniaxial model, the eigenvalues showing the level crossing with field
 can also be obtained from the topological phase in particle wave function occuring
 when a particle completely moves around a unit circle.
 For biaxial spin system with magnetic field along the hard axis, particle wave function
 has the topological phase term due to field in the boundary condition
 as in the uniaxial case, and this phase gives the imaginary part of
 Euclidean action analogous to the Wess-Zumino phase in $(\theta, \phi)$ spin coherent-state representation.
 Therefore, in particle mapping formalism the oscillation of tunnel splitting can be understood as the topological
 effect of a particle moving in a periodic potential, that is, the constrained $S^1$ geometry with the boundary
 condition giving the geometrical phase.

 \newpage
 \vspace{5cm}

\begin{thebibliography}{99}

 \bibitem{chudnovsky88} E. M. Chudnovsky and L. Gunther, {\it Phys. Rev. Lett.} {\bf 60}, 661 (1988);
 B. Barbara and E. M. Chudnovsky, {\it Phys. Lett.} {\bf A145}, 205 (1990).

 \bibitem{gunther95} L. Gunther and B. Barbara, {\it Quantum Tunneling of Magnetization - QTM'94}
 (Kluwar, Dordrecht, 1995) and references therein.

 \bibitem{awschalom90} D. D. Awschalom, M. A. McCord, G. Grinstein, Phys. Rev. Lett. {\bf 65}, 783 (1990);
 D. D. Awschalom, et. al., Science, {\bf 258}, 414 (1992);
 D. D. Awschalom, et. al., Phys. Rev. Lett. {\bf 68}, 3029 (1992);
 S. Gider, D. D. Awschalom, T. Douglas, S. Mann and M. Chaparala, Science, {\bf 268}, 77 (1995).

 \bibitem{friedman96} J. R. Friedman, H. P. Sarachik, J. Tejada and R. Ziolo, Phys. Rev. Lett. {bf 76}, 3830 (1996);
 L. Thomas, F. Lionti, R.Ballou, D. Gatteschi, R. Sessoli and B. Barbara, Nature, {\bf 383}, 145 (1996).
 J. M. Hernandez, X. X. Zhang, F. Luis, J. Bartolome, J. Tejada and R. Ziolo, Europhys. Lett. {\bf 35}, 301 (1996);

 \bibitem{sangregorio97} C. Sangregorio, T. Ohm, C. Paulsen, R. Sessoli and D. Gatteschi, Phys. Rev. Lett. {\bf 78},
 4645 (1997).

 \bibitem{liang98} J. Q. Liang, H. J. W. M\"{u}ller-Kirsten, D. K. Park and F. Zimmerschied, Phys. Rev. Lett.
 {bf 81}, 216 (1998).

 \bibitem{scharf87} G. Scharf, W. F. Wreszinski and J. L. van Hemmen, J. Phys. A: Math. Gen. {\bf 20}, 4309 (1987);
 O. B. Zaslavskii, Phys. Lett. {\bf A145}, 471 (1990).

 \bibitem{ulyanov92} V. V. Ulyanov and  O. B. Zaslavskii, Phys. Rep. {\bf 216}, 179 (1992).

 \bibitem{chudnovsky97} E. M. Chudnovsky and D. A. Garanin, Phys. Rev. Lett. {\bf 79}, 4469 (1997);
 D. A. Garanin and E. M. Chudnovsky, Phys. Rev. B {\bf 56}, 11,102 (1997).

 \bibitem{park99} C. S. Park, S. K. Yoo, D. K. Park and D. H. Yoon, Phys. Rev. B {\bf 59}, 13,581 (1999).

 \bibitem{loss92} D. Loss, D. P. DiVincenzo and G. Grinstein, Phys. Rev. Lett. {\bf 69}, 3232 (1992).

 \bibitem{delft92} J. von Delft and C. L. Henley, Phys. Rev. Lett. {\bf 69}, 3236 (1992).

 \bibitem{fradkin91} E. Fradkin, {\it Field Theories of Condensed Matter Systems} (Addison-Wesley, New York, 1991).

 \bibitem{garg93} A. Garg, Europhys. Lett. {\bf 22}, 205 (1993).

 \bibitem{wernsdorfer99} W. Wernsdorfer and R. Sessoli, Science {\bf 284}, 133 (1999).

 \bibitem{chudnovsky99}
 S. P. Kou, J. Q. Liang, Y. B. Zhang and F. C. Pu, Phys. Rev. B {\bf 59}, 11,792 (1999);
 A. Garg,  Phys. Rev. B {\bf 60}, 6705 (1999); E. M. Chudnovsky and X. M. Hidalgo, cond-mat/9902218.

 \bibitem{add1} Actually, the effective potential, Eq. (\protect{\ref{bipot1}}) shows completely different physics
 from the original system. While the Hamiltonian, Eq.(\protect{\ref{biham1}}) shows the tunneling
 between two equivalent minima for the entire range of $0<h<1$, in the effective particle system,
 a particle decays from the metastable to stable states for $0<h<b$, there is no tunneling at all in the range of $h>b$.
 We only obtain the information of inverted eigenvalues from the periodic potentials and boundary condition
 of wave function.

 \end{thebibliography}
 \end{document}